# The Conception of Thermonuclear Reactor on the Principle of Gravitational Confinement of Dense High-temperature Plasma

Stanislav Fisenko, Igor Fisenko

"Rusthermosinthes" JSC,

6 Gasheka Str., Ducat Place 3, office 1210

Moscow 125047, Russia

Tel: +7 495 956-82-46, Fax: +7 495 956-82-47

E-mail: StanislavFisenko@yandex.ru

**Abstract**

The work of Fisenko S. I., & Fisenko I. S. (2009). *The old and new concepts of physics, 6* (4), 495, shows the key fact of the existence of gravitational radiation as a radiation of the same level as electromagnetic. The obtained results strictly correspond to the framework of relativistic theory of gravitation and quantum mechanics. The given work contributes into further elaboration of the findings considering their application to dense high-temperature plasma of multiple-charge ions. This is due to quantitative character of electron gravitational emission spectrum such that amplification of gravitational emission may take place only in multiple-charge ion high-temperature plasma.

**Introduction**

Last years' astronomical observations have brought to general relativity based astrophysics and cosmology such notions as "inflation", "dark matter", and "dark energy" thus urging to elaboration of the major number of recent alternatives to GR. New theories offer interpretation of these experimental data not invoking those notions for they seem to be wrong or artificial to the authors of these theories. The basic concept implies that gravity must agree with GR at least within Solar System at present epoch but may be essentially different on galaxy scale or higher as well as in early Universe. However all experimental attempts to detect gravitational radiation (based both on GR views and on alternative theories) yield no results. In elaboration of the relativistic theory of gravitation (namely the relativistic theory of gravitation rather than its particular case such as GR) the authors have obtained a model of gravitational interaction at quantum level having no equivalents and making gravitational radiation spectrum computing possible. These results are particularized in the work of Fisenko S. I. and Fisenko I. S. (2009). Generalized summary of those results followed by their possible development is described in section 1 hereafter. In applied perspective, the most important consequence of the properties of the radiated gravitational field (as a field of the same level as electromagnetic one) is compression of the radiating system by it. This property is directly relevant to controlled thermonuclear fusion which is discussed in section 2 including layout of the approbated test unit for plasma compression by the radiated gravitational field (Fisenko & Fisenko, 2005).

**1. Gravitational radiation as a radiation of the same level as electromagnetic**

As is known, the generally covariant form of equations of the Einsteinian relativistic gravitation theory is as follows,

$$R_{ik} - \frac{1}{2} g_{ik} R - \Lambda g_{ik} = \chi T_{ik} \qquad (1)$$

In these equations $\chi$ is a constant linking the geometric properties of space-time with physical matter distribution; the origin of the equations is therefore not connected with the numeric limitation of the $\chi$ quantity value. It is only the necessity of correspondence to the classical Newtonian gravitation theory that leads to the numeric values $\Lambda = 0, \chi = 8\pi G / c^4$, where G is the Newtonian gravitation constant. Equations with constants found like this are actually the equations of the Einsteinian general relativity theory (GR). At the same time, the equations (1) offer a general mathematical form for gravitational field equations; the form complies with the equivalence principle and the postulate of general covariance. The equations of the form (1) were found by Hilbert (1915) simultaneously with Einstein and independently of him. The work of Fisenko and Fisenko (2009) makes a simple but strict suggestion that the quantum area may contain numeric values of the constants K and Λ, that account for steady states in proper gravitational field. It is these states that make up the sources of gravitational field with the Newtonian gravitational constant. The numeric values K and Λ are measured independently in the framework of the approach under consideration. In this connection A. Salam's work (Siravam & Sinha, 1979) should be mentioned, as he was one of the first to pay attention to the fact that the numeric value of the Newtonian gravitational constant is not applicable for the quantum level. It was he who suggested the conception of "strong" gravitation based on the assumption that there are f-mesons spin 2 which make up a SU(3)-multiplet (described by Paul-Firz equations). It was proven that the possibility of a different link constant alongside the Newtonian one does not contradict the observed effects (Siravam & Sinha, 1979). Due to a number of reasons this approach was not developed further. As it is clear now, this "strong" gravitation constant is to be used in equations (1) with $\Lambda \neq 0$. Besides, it is with $\Lambda \neq 0$ that stationary solutions to the Einsteinian general equations can be found. This was mentioned by Einstein himself, but after Friedman (1922) discovered non-stationary solutions with $\Lambda = 0$, the GR theory was finally formed as it is known now. In the framework of the GR the key argument to equate the Λ-element to zero is the necessity of correct limit passage to the Newtonian gravitation theory.

In its simplest approximation (from the point of view of the original mathematical estimates) the problem on steady states in proper gravitational field (with constants K and $\Lambda$) is solved by Fisenko and Fisenko (2009). The solution of this problem provides the following conclusions.

a) With the numeric values $K \approx 5.1 \times 10^{31}$ Nm$^2$kg$^{-2}$ and $\Lambda = 4.4 \times 10^{29}$ m$^{-2}$ there is a spectrum of steady states of the electron in proper gravitational field (0.511 MeV …0.681 MeV). The basic state is the observed electron rest energy 0.511 MeV.

b) These steady states are the sources of the gravitational field with the G constant.

c) The transitions to stationary states of the electron in proper gravitational field cause gravitational emission, which is characterized by constant K, i. e. gravitational emission is an emission of the same level as electromagnetic (electric charge e, gravitational charge $m\sqrt{K}$ ). In this respect there is no point in saying that gravitational effects in the quantum area are characterized by the G constant, as this constant belongs only to the macroscopic area and cannot be transferred to the quantum level (which is also evident from the negative results of registration of gravitational waves with the G constant, they do not exist).

d) The spectrum of electron stationary states in proper gravitational field and transitions to stationary states are shown in Fig. 1. It should be immediately mentioned that the numeric value of the spectrum is approximate. The numeric value of the first stationary state is characterized by the highest inaccuracy, while the accuracy of the estimate increases as it comes closer to $E_\infty = 171 keV$. The existence of electron stationary states in proper gravitational field also corresponds fully to special relativity theory. The relativistic connection between the energy and the impulse is broken if we assume that full energy of an electron is determined only by Lorentz electromagnetic energy (Pauli, 1958).

e) Existence of another consequence is possible which certainly is of interest.

A problem in stationary states of elementary source in proper gravitational field to the fine structure approximation specified by relativism reduces to the solution of the following equations:

$$f'' + \left(\frac{v' - \lambda'}{2} + \frac{2}{r}\right)f' + e^\lambda \left(K_n^2 e^{-v} - K_0^2 - \frac{l(l+1)}{r^2}\right)f = 0 \tag{2}$$

$$-e^{-\lambda}\left(\frac{1}{r^2} - \frac{\lambda'}{r}\right) + \frac{1}{r^2} + \Lambda = \beta(2l+1)\left\{f^2\left[e^{-\lambda}K_n^2 + K_0^2 + \frac{l(l+1)}{r^2}\right] + f'^2 e^{-\lambda}\right\} \tag{3}$$

$$-e^{-\lambda}\left(\frac{1}{r^2} + \frac{v'}{r}\right) + \frac{1}{r^2} + \Lambda = \beta(2l+1)\left\{f^2\left[K_0^2 - K_n^2 e^{-v} + \frac{l(l+1)}{r^2}\right] - e^\lambda f'^2\right\} \tag{4}$$

$$\left\{-\frac{1}{2}(v'' + v'^2) - (v' + \lambda')\left(\frac{v'}{4} + \frac{1}{r}\right) + \frac{1}{r^2}(1 + e^\lambda)\right\}_{r=r_n} = 0 \tag{5}$$

$$f(0) = const \ll \infty \tag{6}$$

$$f(r_n) = 0 \tag{7}$$

$$\lambda(0) = v(0) = 0 \tag{8}$$

$$\int_0^{r_n} f^2 r^2 dr = 1 \tag{9}$$

Equations (2)–(4) result from equations (10)–(11)

$$\left\{-g^{\mu\nu}\frac{\partial}{\partial x_\mu}\frac{\partial}{\partial x_\nu} + g^{\mu\nu}\Gamma^\alpha_{\mu\nu}\frac{\partial}{\partial x_\alpha} - K_0^2\right\}\Psi = 0 \tag{10}$$

$$R_{\mu\nu} - \frac{1}{2}g_{\mu\nu}R = -\kappa\left(T_{\mu\nu} - \mu g_{\mu\nu}\right), \tag{11}$$

on substitution in them for $\Psi$ of the form $\Psi = f_{El}(r)Y_{lm}(\theta, \varphi)\exp\left(\frac{-iEt}{\hbar}\right)$ and specific calculations in the metric of centrally symmetrical field where interval is defined as

$$dS^2 = c^2 e^v dt^2 - r^2(d\theta^2 + \sin^2\theta d\varphi^2) - e^\lambda dr^2 \tag{12}$$

As denoted hereinabove: $f_{El}$ – radial wave function describing states with defined energy $E$ and orbital momentum $l$ ($El$ indices are omitted hereinafter), $Y_{lm}(\theta,\varphi)$ – spherical functions, $K_n = E_n/\hbar c$, $K_0 = cm_0/\hbar$, $\beta = (\kappa/4\pi)(\hbar/m_0)$.

Condition (5) is a condition to find $r_n$. Conditions (6)–(9) are boundary conditions and a normalization requirement for $f$ function, respectively. Condition (5) in the general case is given by $R(K,r_n) = R(G,r_n)$. Neglecting proper gravitational field with constant $G$ this condition may be written as $R(K,r_n) = 0$, which equation (5) corresponds to.

Right hand sides of equations (3)–(4) are computed from the general expression for energy-momentum tensor of complex scalar field:

$$T_{\mu\nu} = \Psi^+_{,\mu}\Psi_{,\nu} + \Psi^+_{,\nu}\Psi_{,\mu} - \left(\Psi^+_{,\mu}\Psi^{,\mu} - K_0^2\Psi^+\Psi\right) \tag{13}$$

The respective components $T_{\mu\nu}$ are deduced by summation over $m$ index using typical identities for spherical functions on substitution of $\Psi = f(r)Y_{lm}(\theta,\varphi)\exp\left(\dfrac{-iEt}{\hbar}\right)$ function into (13).

Energy spectrum estimation given on Fig. 1 is made exactly within this approach. Boundary conditions (6)-(8) accurately correspond to this problem in stationary states as the sates of equilibrium in consequence of interaction with proper gravitational field. A closed trajectory defined mathematically by wave function and set of quantum numbers with orbital momentum values among them corresponds to all the stationary states (beginning with the ground one). We return to estimation of the numerical value of K. Using Kerr-Newman metric (Newman et all, 1965) for estimation of the numerical value of K one can obtain the formula (Fisenko et all, 1990)

$$K = \frac{r^2}{(mcr^2/L - L/mc)(m/rc^2 - e^2/r^2c^4)}; \tag{14}$$

where $r, m, e, L$ are classical electron radius, mass, charge, orbital momentum respectively, and $c$ is the speed of light.

Despite the fact that we used external metric and orbital momentum in deriving the formula (14), its use is legitimate for the orbital momentum of a particle in internal metric equal to the electron spin by an order of magnitude. The estimation of K from the formula (14) using the numerical values of the abovementioned arguments agrees with the estimation that stands in correspondence with numerical values of electron energy spectrum in proper gravitational field. This may suggest that the physical nature of spin is possibly such that these are just values of the orbital momentum of a particle in proper gravitational field.

*Thus:*

f) Closed consistent model of gravitational interaction on quantum level is used that stands in full conformity with quantum mechanics and general equations of the relativistic theory of gravity. A notion of gravitational radiation as a radiation of the same level as electromagnetic one is based on the theoretically proved and experimentally verifiable fact (Fisenko & Fisenko, 2009) of electron stationary states existence in proper gravitational field characterized by gravitational constant $K=10^{42}G$ (G is the Newtonian gravitation constant).

g) Such properties of emitted gravitational field are of immediate applied interest. In particular, compression of emitting system when gravitational emission is amplified is directly relevant to the problem of obtainment of stable states of high-temperature dense plasma for nuclear fusion purposes.

## 2. Physical layout of plasma confinement by the emitted gravitational field type fusion reactor

The problem of controlled nuclear fusion implementation is unequivocally linked to attaining stable states of dense high-temperature plasma. Basic existing results of this problem solving are as follows.

From the state of the art a heavy-current pulsed discharge is known, which is shaped with the aid of a cylindrical discharge chamber (whose end faces function as electrodes) which is filled with a working gas (deuterium, hydrogen, a deuterium-tritium mixture at a pressure of 0.5 to 10 mm Hg, or noble gases at a pressure of 0.01 to 0.1 mm Hg). Then a discharge of a powerful capacitor battery is effected through the gas, with the voltage of 20 to 40 kV supplied to the anode and the current in the forming discharge reaching about 1 MA. In experiments (Lukyanov, 1975) first the first phase of the process was observed — plasma compression to the axis by the current magnetic field with decrease of the current channel diameter by approximately a factor of 10 and formation a brightly glowing plasma column on the discharge axis (z-pinch). In the second phase of the process a rapid development of current channel instabilities (kinks, helical disturbances, etc.) were observed.

The buildup of these instabilities occurs very rapidly and leads to the degradation of the plasma column (plasma jets outburst, discharge discontinuities, etc.), so that the discharge lifetime is limited to a value on the order of $10^{-6}$ s. For this reason in a linear pinch it turns out to be unreal to fulfill the conditions of nuclear fusion defined by the Lawson criterion $n\tau > 10^{14}$ cm-3s, where n is the plasma concentration, $\tau$ is the discharge lifetime..

A similar situation takes place in a $\Theta$-pinch, when to a cylindrical discharge chamber an external longitudinal magnetic field inducing an azimuthal current is impressed.

Magnetic traps are known, which are capable of confining a high-temperature plasma for a long time (but not sufficient for nuclear fusion to proceed) within a limited volume (Artsimovich, 1969). There exist two main varieties of magnetic traps: closed and open ones.

Magnetic traps are devices which are capable of confining a high-temperature plasma for a sufficiently long time within a limited volume and which are described by Artsimovich (1969).

To magnetic traps of closed type (on which hopes to realize the conditions of controlled nuclear fusion (CNU) were pinned for a long time) there belong devices of the Tokamak, Spheromak and Stellarator type in various modifications (Lukyanov, 1975).

In devices of the Tokamak type a ring current creating a rotary transformation of magnetic lines of force is excited in the very plasma. Spheromak represents a compact torus with a toroidal magnetic field inside a plasma. Rotary transformation of magnetic lines of force, effected without exciting a toroidal current in plasma, is realized in Stellarators (Volkov et al., 1983).

Open-type magnetic traps with a linear geometry are: a magnetic bottle, an ambipolar trap, a gas-dynamic type trap (Ryutov, 1988).

In spite of all the design differences of the open-type and closed-type traps, they are based on one principle: attaining hydrostatic equilibrium states of plasma in a magnetic field through the equality of the gas-kinetic plasma pressure and of the magnetic field pressure at the external boundary of plasma. The very diversity of these traps stems from the absence of positive results.

When using a plasma focus device (PF) (this is how an electric discharge is called), a non-stationary bunch of a dense high-temperature (as a rule, deuterium) plasma is obtained (this bunch is also called "plasma focus"). PF belongs to the category of pinches and is formed in the area of current sheath cumulation on the axis of a discharge chamber having a special design. As a result, in contradistinction to a direct pinch, plasma focus acquires a non-cylindrical shape (Petrov et al., 1958).

Unlike linear pinch devices, where the function of electrodes is performed by the chamber end faces, in the PF the role of the cathode is played by the chamber body, as a result of which the plasma bunch acquires the form of a funnel (thence the name of the device). With the same working parameters as in the cylindrical pinch, in a PF device a plasma having higher temperature, density and longer lifetime is obtainable, but the subsequent development of the instability destroys the discharge, as is the case in the linear pinch (Burtsev et al., 1981), and stable states of plasma are actually not attained.

Non-stationary bunches of high-temperature plasma are also obtained in gas-discharge chambers with a coaxial arrangement of electrodes (using devices with coaxial plasma injectors). The first device of such kind was commissioned in 1961 by J. Mather (1961). This device was developed further (in particular, see Brzosko et al., 1994; Brzosko et al., 1991). An essential element of this development was the use of a working gas doped with multielectron atoms. Injection of plasma in such devices is attained owing to the coaxial arrangement of cylindrical chambers, wherein the internal chamber functioning as the anode is disposed geometrically lower than the external cylinder — the cathode. In the works of J. Brzosko it was pointed out that the efficiency of the generation of plasma bunches increases when hydrogen is doped with multielectron atoms. However, in these devices the development of the instability substantially limits the plasma lifetime as well. As a result, this lifetime is smaller than necessary for attaining the conditions for a stable course of the nuclear fusion reaction. With definite design features, in particular, with the use of conical coaxial electrodes (Kozlov & Morozov, 1984), such devices are already plasma injection devices. In the above-indicated devices (devices with coaxial cylindrical electrodes) plasma in all the stages up to the plasma decay, remains in the magnetic field area, though injection of plasma into the interelectrode space takes place. In pure form injection of plasma from the interelectrode space is observed in devices with conical coaxial electrodes. The field of application of plasma injectors is regarded to be auxiliary for plasma injection with subsequent use thereof (for example,. for additional pumping of power in devices of Tokamak type, in laser devices, etc.), which, in turn, has limited the use of these devices not in the pulsed mode, but in the quasi-stationary mode.

Thus, the existing state of the art, based on plasma confinement by a magnetic field, does not solve the problem of confining a dense high-temperature plasma during a period of time required for nuclear fusion reactions to proceed, but effectively solves the problem of heating plasma to a state in which these reactions can proceed.

Contribution of the principle of plasma confinement by the emitted gravitational field to existing results of plasma heating consists in the following.

The fundamental characteristic of gravitational radiation is that its spectrum is banded, corresponding to the spectrum of stationary states of the electron in proper gravitational field.

The existence of cascade transitions from the upper excited levels to the lower ones will make electrons that are excited in the keV energy area radiate in the eV area, i. e. this will cause energy transfer down the spectrum to the low-frequency area. This energy transfer mechanism may take place only in case of quenching of spontaneous emission on the lower energy levels of the electron in proper gravitational field, which would eliminate emission with quantum energy in the keV area. A detailed description of the mechanism of the energy transfer down the spectrum will in prospect yield its precise numeric characteristics. Nevertheless, one cannot doubt the mere fact of its existence predetermined by the banded nature of the deceleration gravitational emission spectrum. The low-frequency nature of the deceleration gravitational emission spectrum will bring about its intensification in plasma by virtue of realization of blockage condition $\omega_g \leq 0,5\sqrt{10^3 n_e}$.

The addition of multiple-charge ions into hydrogen plasma (provided the electron shell energy levels of such ions are close to those of electrons in proper gravitational field, see Fig. 2) will result in deceleration of keV level excited states. The deceleration of low excited electron states will be especially effective with a resonance between the excited electron energy and the excitation energy in the ion. It is well known that carbon, krypton and xenon ions (especially the latter) are strong emitting systems, which cause energy losses in plasma under its compression and thermolization. This will really be a drawback on the initial stage of plasma compression and thermolization. But this will also make up an advantage on the final stage, as, with plasma confined by the emitted gravitational field and with a fusion reaction, the energy withdrawal from plasma will take place only in the form of electromagnetic energy emission. Besides, a significant part of this emission will be shifted into optical band exactly under the conditions of fusion reaction.

Coupled with cascade transitions this will lead to gravitational emission energy transfer to the long-wave eV energy area. Thus, the presence of multielectron atoms in hydrogen plasma will provide blockage and intensification of gravitational emission in plasma. It should be mentioned that this scheme is well known in the optical band (fluorescence) and is widely applied for quenching the emission lines of the optical band. This is a mechanism of plasma confinement in the radiated gravitational field, supplementary to the magnetic way of plasma confinement. The above said is especially well pronounced in the equation of particle motion in classical approximation; it is to this approximation that the long-wave area of radiated gravitational field spectrum corresponds. The equation takes the form:

$$\frac{d^2 x^i}{ds^2} + \Gamma^i_{kl} \frac{dx^k}{ds} \frac{dx^l}{ds} = 0 \qquad (15)$$

where $\frac{d^2 x^i}{ds^2}$ is 4-acceleration of the particle and the value $m\Gamma^i_{kl} \frac{dx^k}{ds} \frac{dx^l}{ds}$ is "4-force".

Both mechanisms are obviously to alternate, as it is described in the work of Fisenko and Fisenko (2009): the thermolization of plasma under compression by magnetic field is to be followed by its injection from the magnetic field area to provide the realization of the final stage of plasma compression and confinement by the radiated gravitational field. Thus, gravitational emission may be generated in dense high-temperature plasma and can be intensified under certain conditions; its intensification leads, however, to compression of the radiating system. The succession of actions aimed at obtaining steady states of dense high-temperature plasma is as follows.

- Formation and acceleration of binary plasma with multiple-charge ions by accelerating magnetic field in pulse high-current discharge.
- Binary plasma injection from the accelerating magnetic field area:

excitation of those steady states of the electron in proper gravitational field which lie in the energy range up to 171 keV with further emission. This is to be done under conditions of quenching of lower excited electron energy levels on the energy levels of electron shells of heavy component ions (including quenching electron excited states directly on nuclei with low atomic numbers, such as carbon) in the process of deceleration of plasma bunch injected from the accelerating magnetic field area. Cascade transitions from the upper energy levels provide gravitational emission energy transfer to the long-wave area.

The conditions for such succession of actions are created in the two-section MAGO chamber (see Fig. 3). This structural design (Morhov, Tchernyshov et al., 1979) complies most fully with the suggested way of formation of dense high-temperature plasma steady states (Fisenko & Fisenko, 2005) with magneto-hydrodynamic outflow of plasma and further plasma bunch energy conversion (in the process of deceleration) into plasma thermal energy. This provides further thermolization of plasma, as well as generation of gravitational emission and its transition down the spectrum to the long-wave area with further plasma compression under the conditions of emission blockage and intensification.

Available experimental data show that they can be reproduced in an active experiment directly with thermonuclear plasma on the ground of existence of gravitational radiation narrow-band spectrum in the range up to 171 keV with long-wave spectrum part realized by cascade electron transitions from the upper energy levels. Quenching lower excited states of electrons on the electron shell energy levels of heavy component ions in combination with cascade transitions will result in plasma compression under the conditions of blockage and gravitational radiation intensification.

Operation capacity of the chamber according to the scheme shown in Fig. 3 with deuterium-tritium composition was tested experimentally(Fig.4). The obtained experimental data of plasma compression in the chamber is to be found in the article of Lindemuth et al. (1995). However, the holding time is not sufficient and needs to be longer. The choice of such design for a thermonuclear reactor is unequivocal since it completely corresponds to the system of exciting and amplifying gravitational radiation when plasma is thermolized after the outflow from the nozzle. The required additional compression actually takes place when the working plasma composition is changed (multielectronic ions) and current-voltage characteristic of the charge changes correspondingly.

The simplicity of the chamber technical design is increased by the possibility of its application as an electric load generator, a capacitor bank, as well as an autonomous magnetic explosion generator, with all ensuing practical applications of this thermonuclear reactor.

**Conclusion**

1. A concept of nuclear fusion reactor is offered based on a method for forming stable states of plasma on the basis of confinement by the emitted gravitational field as a field of the same level as electromagnetic.

2. A feasibility of the offered concept is substantiated by consistent though not complete data fitting an integral structure.

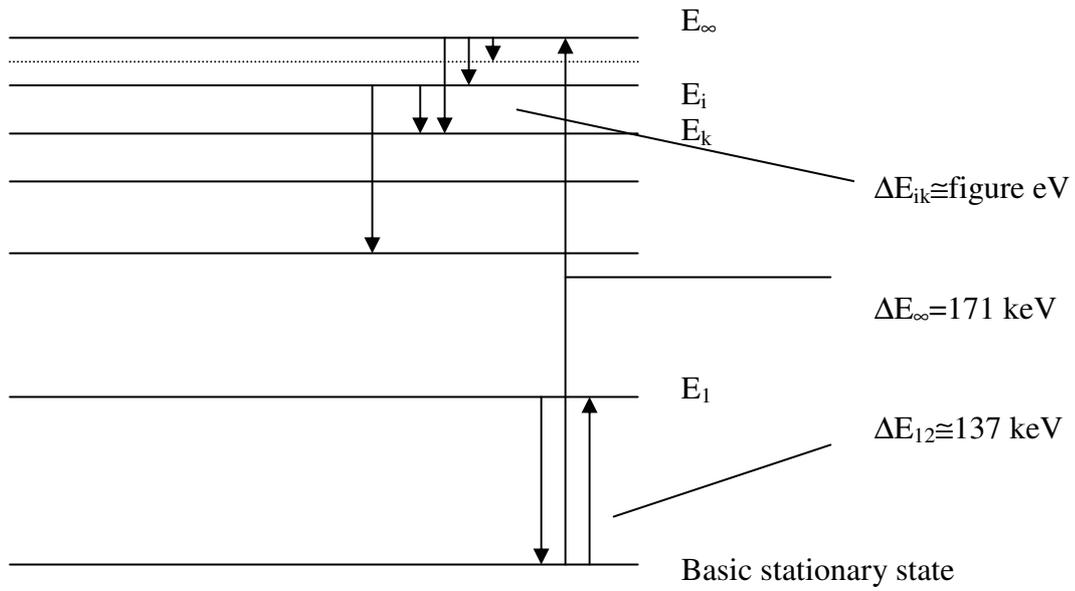

Figure 1. Transitions to stationary states of electrons in proper gravitational field.

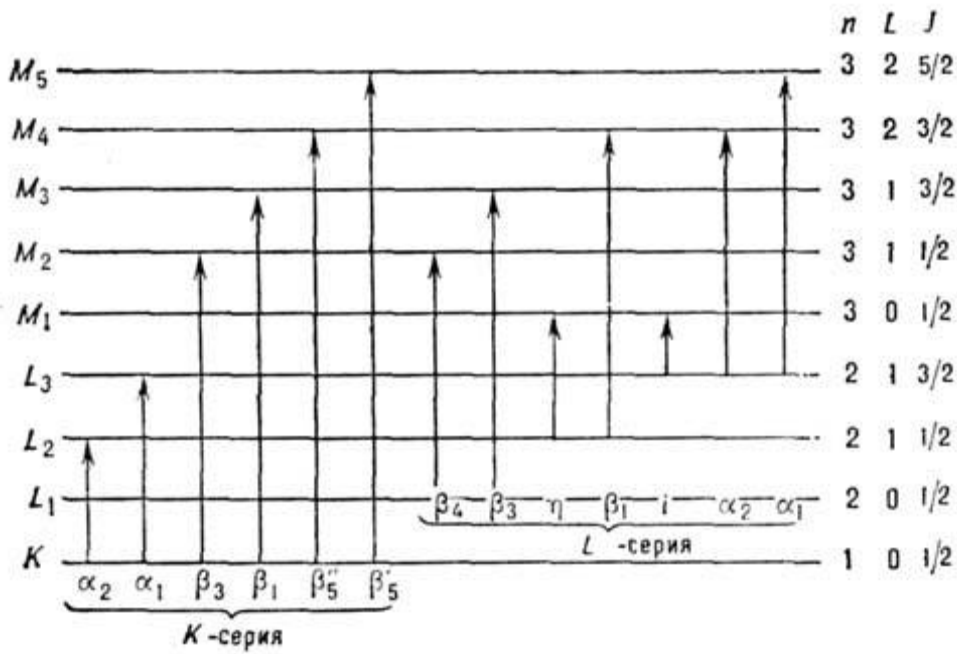

Figuer 2. A scheme of K-, L- and M- energy levels of the atom, and the main lines of K- and L- series; n, l, j are the principal, the orbital and the inner quantum numbers of energy levels к, L1, L2 etc. The energies of photons of the main lines reach units and scores of keV.

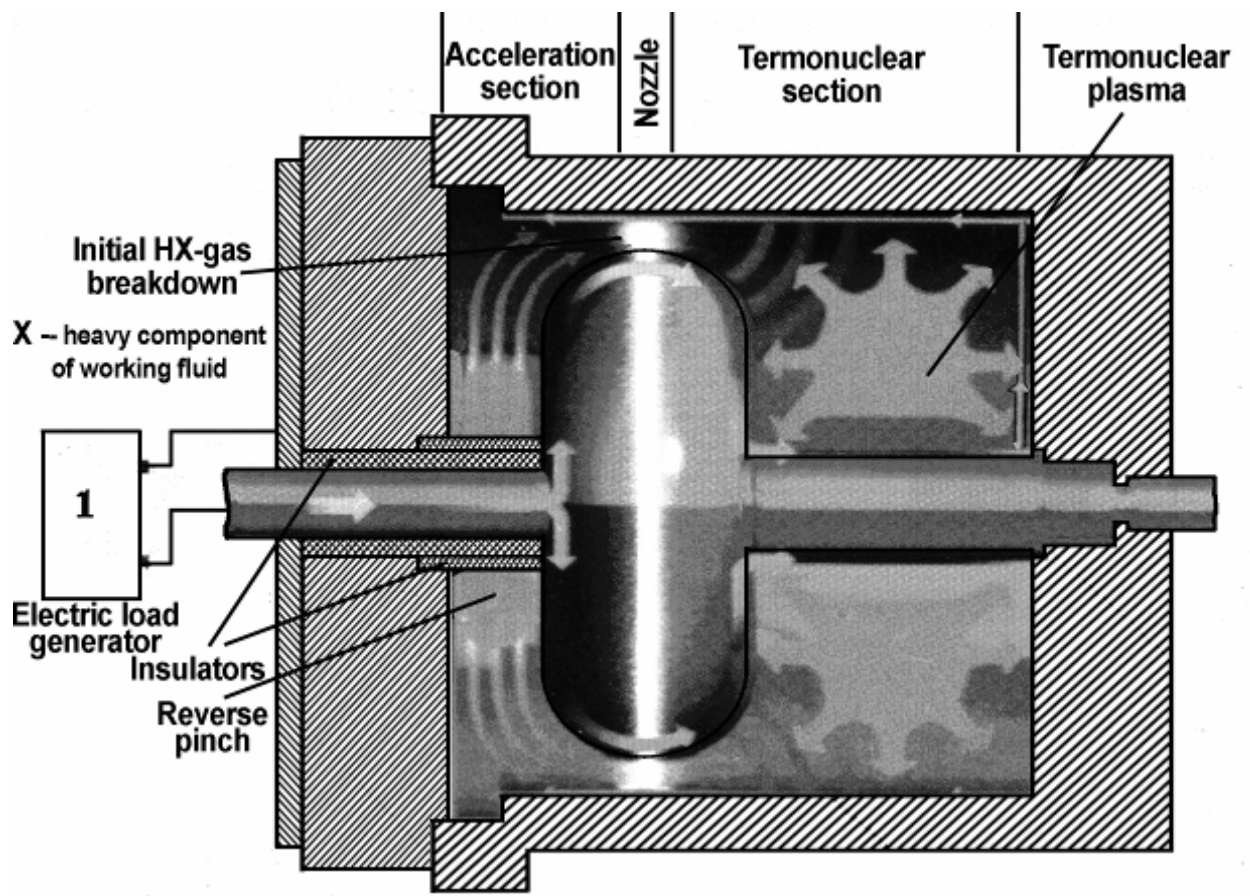

Figure 3. Physical diagram of thermonuclear plasma in MAGO chamber.

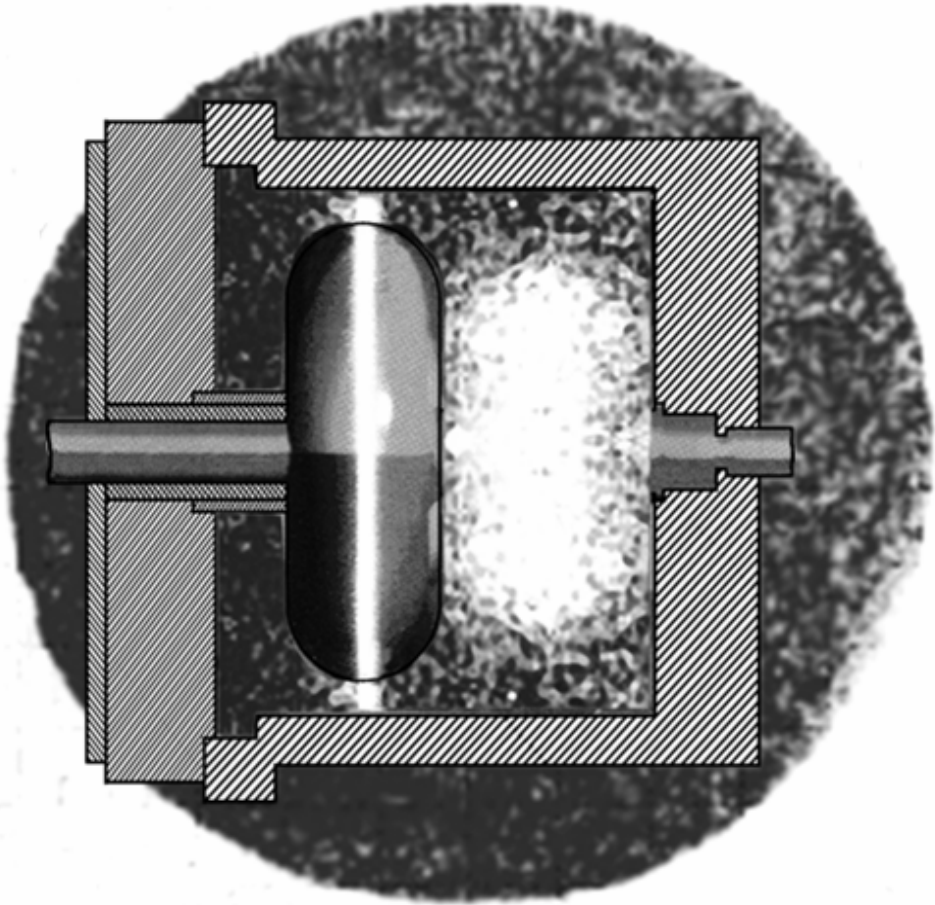

Figure 4. Neutron picture of the neutron generation area.